\newcommand{\spt}[1]{$^{\mathrm{#1}}$}
\newcommand{\sbt}[1]{$_{\mathrm{#1}}$}

\documentclass[aip,reprint]{revtex4-1}

\usepackage{amsmath}
\usepackage{amssymb}
\usepackage{gensymb}
\usepackage{graphicx}
\usepackage{dcolumn}
\usepackage{verbatim}
\usepackage{comment}

\newcounter{tmp}
\newcounter{reaction}
\newenvironment{reaction}{
\setcounter{tmp}{\value{equation}}
\setcounter{equation}{\value{reaction}}

\begin{equation}
}{
\end{equation}
\setcounter{reaction}{\value{equation}}
\setcounter{equation}{\value{tmp}}
}

\begin{document}
	\title{Initial estimate for minimum energy pathways and transition states using velocities in internal coordinates}

	\author{Mark C. Palenik}
	\email{mark.palenik@nrl.navy.mil}
	\address{U.S. Naval Research Laboratory, Chemistry Division, Washington, DC 20375, United States}
	
	\begin{abstract}
		Many algorithms for finding reaction pathways require an initial estimate of the minimum energy path (MEP).  Most estimation methods use a variational approach and thus must be seeded from an even simpler path, such as one generated by Cartesian interpolation.  Often, care must be taken to avoid atomic intersections in this seed path, and the estimator itself may potentially converge to multiple undesirable local minima.  As an alternative we form an initial estimate by numerically integrating a  velocity vector field that is projected from redundant internal coordinates into the Cartesian manifold.  We compare this method to the image dependent pair potential, the geodesic method, and linear Cartesian interpolation using three test cases: the rotation of a methyl group in ethane, HCN$\to$HNC tautomerization, and HONO elimination from dimethylnitramine.  In the first test case, a zero-temperature string calculation seeded with our method converges to the MEP in significantly fewer geometry and SCF cycles than any of the others, while in the second, only the geodesic method slightly outperformed ours.  In the third test case, we used the midpoint of each path as an initial guess for a transition state calculation.  Our midpoint was geometrically the closest to the true transition state and converged in the fewest geometry and SCF cycles.
	\end{abstract}
	
	\maketitle

\section{Introduction}
	
	The minimum energy pathway (MEP) between reactants and products represents the ``path of least resistance'' over which a reaction can occur \cite{Dunitz1975} and therefore defines a reaction coordinate.  It can be defined mathematically in several ways, including from eigenvectors of the Hessian of potential energy with respect to nuclear coordinates and from a minimal action principle. \cite{Tachibana1979,Fukui1981} In transition state theory, the highest point on the MEP, known as the transition state, is used to calculate reaction rates. \cite{Eyring1935,Wigner1937,Truhlar1996}
	
	Computing the MEP requires finding a continuum of points, or an approximation thereof, on a complicated energy surface.  Double-ended computational methods, like the nudged elastic band \cite{Mills1994,Mills1995} and zero-temperature string \cite{Weinan2002,Weinan2007} (ZTS) methods, have to be seeded with an initial guess for the reaction pathway that connects the reactant and product geometries.  Starting from this seed path, a selection of points is refined to meet some optimization condition.  The seed path is generated in a manner that is computationally inexpensive compared to the refinement technique, but the better the initial guess, the faster the refinement process can take place and the more likely it is to converge to the desired result.  A good initial guess for the MEP can also be used to obtain an estimate for the transition state, which can then be optimized to a saddle-point.
	
	Linearly interpolating the Cartesian positions between reactant and product geometries is not usually sufficient to generate the initial guess, because it can lead to atomic intersections and otherwise unrealistic geometries.  However, Cartesian interpolation can sometimes be used as a second seed path for various superior estimation methods that have been developed, although it may still be necessary to avoid atomic intersections, and the choice of seed path can sometimes affect the final outcome of the estimation method.
	
One of the oldest estimation methods, known as linear synchronous transit \cite{Halgreen1977} (LST), improves over Cartesian interpolation by using a set of internal coordinates derived from bond lengths as a heuristic for the reaction coordinate.  Because these internal coordinates are not isomorphic to Cartesian positions, interpolation through them from the reactant to product structures cannot be carried out directly.  For example, unphysical bond lengths may arise that violate the triangle inequality.  LST uses least squares minimization to find the set of Cartesian positions that most closely match the linearly interpolated internal coordinates, but this procedure tends to produce discontinuous pathways and can be prone to outright failure.
	
	Other estimation methods, such as the image dependent pair potential \cite{Smidstrup2014} (IDPP) are likewise based on finding a set of Cartesian positions that closely match some desired set of internal coordinates.  A paper by Zhu et al. \cite{Zhu2019} takes a slightly different approach and treats the Cartesian coordinates as an embedded surface in the space of internal coordinates.  The mapping from Cartesians to internal coordinates then induces a non-Euclidean metric on this surface, and the MEP is estimated as a geodesic between reactants and products.
	
One thing that all of these methods have in common is that they use a variational approach where a seed path is used as input and points along the path are varied to minimize some quantity.  In the geodesic method, for example, a series of paths are generated with a combination LST and random noise, and the one with the shortest length in internal coordinates is used as a seed.  Points along this path are then varied to minimize the total path length, resulting in a geodesic.
		
	Ideally, however, we would like to avoid the need to form another estimated path to seed our estimate.  This can add an additional layer of complication if, for example, care must be taken to avoid atomic intersections in the initial seed path, or if a bad choice of seed path can prevent the estimator from converging.  Furthermore, it is often advantages to keep the implementation of an estimation technique as simple as possible for the user.  A variational estimator may potentially converge to different local minima depending on how it is seeded, and if the seed path for the estimator is generated automatically, it can be difficult to know whether it results in the best possible estimate.
	
	Potentially, one might consider integrating the geodesic equation, rather than using a variational principle to avoid the need for a seed path.  However, the geodesic equation is a second-order differential equation involving derivatives of the metric tensor, which form a complicated three-index tensor.  As a second-order differential equation, it can be written as two coupled first-order differential equations, but the solution then depends on the specification of twice as many initial conditions, namely, initial positions and initial velocities for every atom.  The relationship between the initial velocity and terminal point cannot be determined until the equation has already been integrated, making it exceedingly difficult to define a path that terminates at the correct products.  This also does not solve the problem that, depending on the geometry of the space, the geodesic connecting two points is not necessarily unique.
	
	To that end, we have developed a first-order geometric method where we also treat the Cartesian manifold as an embedded surface in redundant internal coordinates.  We begin by defining a velocity vector field at every point in internal coordinates that points toward the desired product geometry.  We then, through a series of steps, project this vector field into the Cartesian manifold.  Because velocity is the first-derivative of position, this defines a first-order differential equation for a path from any initial geometry that ends at the desired product geometry.  Unlike the geodesic equation, this equation only involves two-index tensors (matrices), and can be numerically integrated without starting from a seed path.  By arc-length parameterizing the path in terms of internal, rather than Cartesian distances, we can also obtain an estimate for the transition state from the midpoint, in much the same way as in LST.
	
	The computation time of our velocity-based method is cubic in the number of atoms, $N$, and linear in the number of points on the path, $P$, or $O(PN^3)$.  While the computation time of any estimation method is trivial in comparison to finding the true MEP, our tests of the geodesic optimization routine described by Zhu et al. showed that it appears to run cubically in the number of points as $O(P^3N^3)$, and runs orders of magnitude more slowly than our method.  We do believe, however, that the scaling could potentially be reduced to $O(PN^3)$ using a modified Newton-Raphson or quasi-Newton optimization on the  entire path that exploits the block-tridiagonal structure of the Hessian, which we describe in more detail in our results under subsection~\ref{SubsecHONO}.
	
	We tested our method in three different systems: the 120\spt{o} rotation of a methyl group in ethane, HCN$\to$HNC tautomerization, and HONO elimination from dimethylnitramine.  For the first two systems, we compared the rate of convergence of ZTS calculations seeded with our method, the geodesic method of Zhu, IDPP, and linear Cartesian interpolation.  The ZTS path for ethane rotation seeded with our method converged the fastest, in only 3 geometry cycles and 173 SCF cycles.  The next best method, the geodesic method, required 14 geometry cycles and 440 SCF cycles, and Cartesian interpolation required 67 geometry cycles and 2086 SCF cycles.  For HCN$\to$HNC tautomerization, our method was second best, but required only an additional 3 geometry cycles and 6 SCF over the geodesic method, while the IDPP method required an extra 100 SCF cycles, and Cartesian interpolation required nearly 500 more.
	
	In the third system, we were unable to converge a ZTS path from any initial estimate, and instead we generated an intrinsic reaction coordinate \cite{Fukui1981} (IRC) path from the transition state.  We tested the midpoint of our method as an initial guess for a transition state optimization and compared it to LST and Cartesian interpolation, as well as midpoints taken from the geodesic and IDPP paths.  Although the geodesic method and IDPP aren't intended to estimate transition states, they were both an improvement over Cartesian interpolation, although not as good as LST.  The midpoint from our method was geometrically closest to the true transition state and converged the fastest of any of the methods we tested.  The estimate from Cartesian interpolation was particularly poor, with an energy 624~kcal/mol above the true transition state, whereas the energy along our path peaked at only 62~kcal/mol above the true transition state.
	
\section{Background}
	
	The ground state energy of a molecule can be written as a function of the relative positions of its nuclei.  These relative positions can equivalently be described in terms of some set of internal coordinates that do not include translational or rotational degrees of freedom.  Many reaction pathway estimation methods use a set of redundant internal coordinates that are functions solely of interatomic distances, and which are typically greater in number than the actual degrees of freedom in the system (hence the term ``redundant'').
	
	Let us define such a set of internal coordinates $\{q_i\}$ that are each a function of distances between a unique pair of atoms.  If we have $N$ atoms then there are $N(N-1)/2$ possible internal coordinates of this type.  The Cartesian positions of a collection of $N$ atoms can be described by $3N$ coordinates, and therefore, we note that only if $N=7$ are the number of Cartesian coordinates and internal coordinates equal.  If $N>7$, there are more internal coordinates than Cartesian positions.
	
	Because the energy of a molecule is invariant under translations and rotations of the entire system, we can remove those degrees of freedom and, if the atoms are allowed to be in any arbitrary configuration (i.e. not confined to a line or plane), specify its geometry with $3N-6$ Cartesian coordinates, given by $\{x^i\}$.  There are more internal coordinates than Cartesian coordinates for $N>4$, which indicates that the mapping $\{x^i\}\to\{q^i\}$ cannot be inverted.  In fact, this is easy to show, even for $N=3$, when the number of internal and Cartesian coordinates are the same.
	
	If the mapping $\{x^i\}\to\{q^i\}$ is invertible, then it must be an isomorphism, meaning every point in $\{q^i\}$ corresponds to a unique point in $\{x^i\}$ and vice versa.  However, this is not the case.  As a simple example, consider three atomic nuclei, $A$, $B$, and $X$.  According to the triangle inequality, there is no set of Cartesian coordinates where the bond length $A-B$ is greater than the sum of $A-C$ and $B-C$.  But while this doesn't define a valid set of cartesian positions in $\{x^i\}$, it does define a perfectly valid set of coordinates in $\{q_i\}$.  Therefore, $\{x^i\}\to\{q^i\}$ is not an isomorphism and is not invertible.
	
	As an additional practical matter, even when there is a set of Cartesian positions corresponding to the desired internal coordinates, it can be difficult to retrieve these positions, given that the internal coordinates may be complicated functions of bond lengths.  LST attempts to solve both of these problems by minimizing the mean-squared difference between the desired internal coordinates determined by interpolation and the actual coordinates computed from the Cartesian atomic positions.  If the set of internal coordinates for the reactant geometry is given by $\{q^{i0}\}$ and the set of coordinates for the product geometry is given by $\{q^{i1}\}$, then we can define an interpolated geometry as
\begin{equation}
(1-t)q^{i0}+tq^{i1},
\end{equation}
which goes from the reactants to the products on the interval $0\leq t\leq 1$.  The equations for linear synchronous transit on this interval can then be written as
\begin{equation}
\frac{\partial}{\partial x^j} \sum_i\left(q^i-(1-t)q^{i0}-tq^{i1}\right)^2 = 0,
\label{EqLST}
\end{equation}
where $q^i$ is an actual internal coordinate at time $t$ computed from the Cartesian positions.  Given the possibility of local minima, the LST path can depend on the choice of seed path.
	
	Even worse, LST tends to produce discontinuous paths and is prone to outright failure at times.  In Sec.~\ref{SecLSTCompare}, we formulate an expression for a continuous LST path as a differential equation and show that in our test case, it does not connect the reactants to the desired products, and that the final point on at $t=1$ differs significantly from the correct geometry.

	However, despite its shortcomings, we can take a lesson from the mean-squared difference in Eq.~(\ref{EqLST}).  This mean-squared difference is equivalent to a dot product in internal coordinates.  If we define a vector $\vec{v}$ that points from the actual internal coordinates $\vec{q}$ to the desired internal coordinates, $(1-t)\vec{q}^0-t\vec{q}^1$, then Eq.~(\ref{EqLST}) is simply 
\begin{equation}
\frac{\partial}{\partial x_j}\left(\vec{v}\cdot\vec{v}\right)=0,
\end{equation}
where the dot product is taken using a Euclidean metric and represents the distance squared in internal coordinates between the actual and desired structures.

If distances are computed in the space of internal coordinates using a Euclidean metric, this induces a non-Euclidean metric in Cartesian coordinates.  The space of Cartesian coordinates is then like an embedded surface in a higher dimensional Euclidean space, which is also the basis of the geodesic method of Zhu et al.  

Here, however, we take inspiration from LST, while still treating the Cartesian coordinates as an embedded surface.  We would like to be able to linearly interpolate between reactant and product geometries, but this is not possible because the line connecting the two structures in internal coordinates does not usually lie entirely in the Cartesian plane.  Therefore, we take a velocity vector that points from the actual internal coordinates at time $t$ toward the internal coordinates of the product geometry and project it into the Cartesian plane.  By integrating this velocity vector, we produce a continuous estimate for the MEP connecting reactants and products.  We will outline the mathematics of this transformation below.
	
\section{Method}

	Given that the dimensionality of the internal coordinates is greater than that of the Cartesian coordinates for $N>4$ atoms, let us treat the Cartesian manifold as an embedded surface in the space of internal coordinates.  The mapping $\{x^i\}\to\{q^i\}$ can then be thought of as a mapping from a manifold $X$ to another manifold $Q$.
	
	If our current structure has the coordinates $\{q^i\}$ in $Q$ and the product structure that we are trying to move toward has the coordinates $\{q^{i1}\}$, the most natural way to move between the two points is along a straight line.  If this movement takes place over the interval $0\leq t\leq 1$, the velocity along this path is simply
	\begin{equation}
	\frac{d q^i}{dt} = \frac{q^{i1} - q^i}{1-t}.
	\label{EqVelocityQ}
	\end{equation}

However, while this defines a path through internal coordinates as a function of time, it is actually the Cartesian coordinates as a function of time that we need.  Therefore, we need to transform the velocity vector into Cartesian coordinates and integrate it.  Employing the transformation law for vectors, we find
\begin{equation}
\frac{dx^i}{dt} = \frac{\partial x^i}{\partial q^j} \frac{dq^j}{dt},
\end{equation}
where we have used Einstein notation in the above equation, with an implied sum over repeated indices, and shall continue to do so for the remainder of the paper.

Unfortunately, as we pointed out previously, while we can define $\{q^i\}$ as a function of $\{x^i\}$, the inverse function does not usually exist.  Therefore, while we can build the Jacobian matrix, $\partial q^i/\partial x^j$, it cannot be inverted.  Given that the dimensionality of $X$ and $Q$ is not the same, it is not even a square matrix.  While we can push forward vectors from $X$ to $Q$, we cannot do so in the opposite direction.

Because $X$ is a lower dimensional surface embedded in $Q$, an arbitrary vector in $Q$ will not necessarily lie entirely in $X$.  In general, a vector in $Q$ will have a component parallel to the tangent plane of a given point in $X$ and a component orthogonal to it.  If the Cartesian coordinates of our current structure are $\{x^i\}$, then we would like find the component of the velocity in Eq.~(\ref{EqVelocityQ}) that lies in the tangent plane of the Cartesian manifold at this point.

In fact, we can take every point in $X$, and define a velocity vector from its corresponding point in $Q$ that points toward $\{q^{i1}\}$.  If we project this vector into the tangent plane of $X$, the resulting vector field defines a differential equation for a set of continuous paths that connect an arbitrary point in $X$ to the desired product geometry.

To project the velocity in $Q$ into the tangent plane of $X$, we can use a geometric object called a 1-form, which we denote with lower indices.  Employing the transformation law for 1-forms, we get
\begin{equation}
\frac{dx_i}{dt} = \frac{\partial q^j}{\partial x^i}\frac{dq_j}{dt}.
\end{equation}
In Euclidean spaces, vectors transform identically into 1-forms, and therefore, the components $dq^i/dt$ are equal to $dq_i/dt$.  But because the Cartesian manifold is not a Euclidean space in this picture we have the relation
\begin{equation}
\frac{dx_i}{dt} = g_{ij}\frac{\partial x^j}{\partial t},
\end{equation}
where $g_{ij}$ is the metric tensor, given by
\begin{equation}
g_{jk} =\frac{\partial q^i}{\partial x^j}\frac{\partial q_i}{\partial x^k},
\end{equation}
which is a square, invertible matrix (except at coordinate singularities, which we touch on briefly when we look at HCN$\to$HNC tautomerization).

Using the matrix-inverse of $g_{jk}$, denoted $g^{jk}$, we can transform a 1-form into its corresponding vector.  The projection of the vector $dq^i/dt$ into the Cartesian manifold is then
	\begin{equation}
	    \frac{dx^k}{dt} = g^{jk}\frac{\partial q^i}{\partial x^j}\frac{q_i^1-q_i}{1-t}.
	    \label{EqDiffX}
	\end{equation}
	
	Integrating Eq.~(\ref{EqDiffX}) from $t=0$ to $t=1$ takes $x^i$ along a continuous path from the initial structure to the final structure.  However, $x^i$ will not necessarily change at a uniform rate, which can result in an uneven pathway with large changes in geometry between neighboring.  To remedy this, we can reparametrize $x$, using the arc-length parameter $\tau$ in internal coordinates, where
	\begin{equation}
	    \begin{split}
	    \frac{d\tau}{dt} &= \sqrt{\frac{dx^i}{dt}g_{ij}\frac{dx^j}{dt}}\\
	    &=\sqrt{\frac{\partial q^a}{\partial x^i}\frac{\left(q_a^1-q_a\right)g^{ij}\left(q_c^1-q_c\right)}{(1-t)^2}\frac{\partial q^c}{\partial x^j}},
	    \end{split}
	\end{equation}
	leaving us with
	\begin{equation}
	    \frac{dx^k}{d\tau} = g^{jk}\frac{\partial q^i}{\partial x^j}\frac{q_i^1-q_i}{\sqrt{\frac{\partial q^a}{\partial x^l}\left(q_a^1-q_a\right)g^{lm}\left(q_c^1-q_c\right)\frac{\partial q^c}{\partial x^m}}}.
	    \label{EqdxFinal}
	\end{equation}
	This is not the only possible alternative parametrization.  One could potentially, for example, reparametrize in terms of the Cartesian arc-length.  However, this particular parametrization is convenient because the midpoint in internal coordinates is often a good estimate for the transition state.  The LST midpoint, for example, is frequently used in this manner, and in Sec.~\ref{SubsecHONO}, we show an example where the midpoint from our method results in particularly efficient convergence of a saddle-point optimization.
			
	The downside to an arc-length parametrization is that the total length of the path is not known ahead of time, as a straight line through Cartesian coordinates is not the shortest possible path through generalized coordinates.  However, in practice, the integration can be performed until the current coordinates differ from the product structure coordinates by some small amount, at which point, the end of the path has been reached.

\subsection{Choice of coordinates and treatment of singularities}
\label{SecCoordSing}

	Many different choices of internal coordinates are possible, and we have been able, for example, to achieve reasonable-looking reaction pathways by defining our coordinates as the interatomic distance divided by the sum of the covalent radii.  However, we ultimately settled on a basis of coordinates slightly modified from those defined in the paper by Zhu et al. \cite{Zhu2019} given by
	\begin{equation}
	    q_{kl} = \exp{\left[-\alpha\frac{r_{kl}-r_{kl}^e}{r_{kl}^e}\right]}+\beta\frac{r_{kl}^e}{r_{kl}}+\sigma \frac{r_{kl}}{r_{kl}^e},
	    \label{Eqqkl}
	\end{equation}
	where $r_{kl}$ is the distance between atoms $k$ and $l$ and $r^e_{kl}$ is the sum of their covalent radii.  The parameters $\alpha$ and $\beta$ are taken from the same paper and set to 1.7 and 0.01 respectively.  The parameter $\sigma$, which we added, was set to 0.01045.
	
	Because the number of $q_{kl}$ coordinates is typically greater than the number of Cartesian coordinates with translational and rotational degrees of freedom removed, there are a minimum of $N(N-1)/2 - (3N-6) = N^2-7N/2 + 6$ singularities in the Jacobian matrix $\partial q^i/\partial x^j$.  These singularities are not present in the metric tensor and therefore do not cause a problem with the matrix inverse required to obtain $g^{jk}$ from $g_{jk}$.
	
	There are, however, two situations in which singularities may still cause a problem.  First, note that in Eq.~(\ref{EqdxFinal}), the Jacobian is contracted with a vector proportional to $q_i^1 - q_i$.  If this vector points along a singular eigenvector of the Jacobian, it will cause $dx^k/d\tau$ to become zero and the path will prematurely terminate.  Such a situation corresponds to the velocity vector through internal coordinates pointing in a direction orthogonal to the Cartesian manifold.  The second situation in which problems arise is when there are more than $N^2-7N/2+6$ singularities in the Jacobian matrix, and the metric tensor itself becomes singular.
	
	We deal with the first problem by introducing the parameter $\sigma$.  When $\sigma=0$, we found that $q_i^1-q_i$ occasionally becomes orthogonal to the Cartesian manifold in certain noncovalently bonded structures and during rotations, where there is little change in the distance between neighboring atoms.  Likely, this is because the first two terms of $q_{kl}$ rapidly approach zero as the distance between atoms grows, regardless of their relative configurations.  Empirically, we have found that the small value of sigma we picked fixes this problem.
	
	Because the dynamics of the path are dependent on derivatives of the coordinates with respect to Cartesian positions, the effect of the $\sigma$ term is independent of the distance between atoms.  Therefore, when the atoms are close together and the derivatives of the $\alpha$ and $\beta$ terms are large, these two terms will dominate the behavior of the pathway.  When the distance between atoms is large, the $\sigma$ term will dominate.  If some distances are small and some distance are large, the dominant effect on a given atom is determined by the relative values of $\alpha$, $\beta$, and $\sigma$.
	
	We would like $\sigma$ to be small enough that it does not affect the dynamics of atoms that are close enough together to interact strongly, but large enough that it is not swamped by the $\alpha$ and $\beta$ terms of neighboring atoms when it is necessary.  The value of $\sigma=0.01045$ was determined by finding the smallest value that would allow us to compute the pathway for a 120\spt{o} rotation of a methyl group in ethane.
	
The second problem, relating to singularities of the metric tensor, occurs at high-symmetry points, when the reaction pathway must break the symmetry of a molecule but there are multiple equivalent ways of doing so.  We give an example of such a situation in Sec.~\ref{SecHCN} with HCN$\to$HNC tautomerization, where the two endpoints of the path are linear structures and the intermediate points are not.

We deal with this problem by diagonalizing $g_{jk}$ and modifying both its nearly singular eigenvalues, as well as the components of the velocity 1-form that point along those directions.  In the basis of eigenvectors, the vector-matrix multiplication $g^{jk}V^j$ becomes $V'_k/\epsilon_k$ (with no summation over $k$), where $V'$ is related to $V$ by a unitary transformation.

If $\epsilon_k$ is zero, this obviously causes a problem, and a common approach in, for example, least squares fitting, is to simply throw out the singular values.  However, singular values here are typically indicative of high symmetry points, and, even if the reactants and products have the same high symmetry, it usually needs to break along the reaction pathway.  This must be done artificially, and we do so by giving the system some arbitrary velocity along a symmetry breaking direction.

If $\varepsilon_k=0$, we can break the symmetry simply by setting $\varepsilon=1$ and $V'_k=1$.  Likewise, if $|\varepsilon_k|$ is above some threshold $\Delta$, we do not modify $\varepsilon_k$ or $V'$.  In the intermediate range, where $0\leq|\varepsilon_k|\leq\Delta$, we linearly interpolate between those two behaviors, so that we have
\begin{equation}
\frac{V'_k}{\epsilon_k}\to \frac{|\epsilon_k|}{\Delta}\left(\frac{V'_k}{\Delta}-1\right)+1.
\end{equation}
In our calculations, we set $\Delta=1\times 10^{-12}$.

The reason we use linear interpolation rather than changing the behavior abruptly when $|\epsilon_k|\geq\Delta$ is that in some systems, the symmetry is not actually supposed to break.  Consider, for example, a hydrogen atom being transferred along a straight line between two other atoms.  When all three atoms lie along a line, the metric tensor is singular, and our equation drives the hydrogen in an arbitrary direction away from symmetry.  Once the symmetry is broken, however, and $g_{jk}$ is no longer singular, Eq.~(\ref{EqdxFinal}) actually drives the system in the opposite direction back toward the high symmetry point.  The abrupt oscillation back and forth between these two behaviors causes problems for the integrator and the integration fails to complete.  In practice, we have found that using linear interpolation results in a point where the two opposing forces balance, the symmetry is imperceptibly broken.  While the integration is slow in such situations, it eventually results in a reasonable path.

\section{Results}

We tested our velocity-based method on three different systems and compared it to the geodesic method of Zhu et al., the IDPP method, and linear Cartesian interpolation.  For the first two systems, methyl group rotation in ethane and HCN$\to$HNC tautomerization, we computed 11 point paths with all four methods and refined them with a ZTS calculation in NWChem\cite{Valiev2010}.  The ZTS calculations were performed with three different step sizes, 0.1, 0.5, and 1.0, and results were reported for the step size that led to the fastest convergence with each method.  The number of geometry and SCF cycles required to complete each calculation, along with the step size that lead to the fastest convergence, are given in Table~\ref{TableCompare}. 

The pathway for our velocity-based method was computed by integrating Eq.~(\ref{EqdxFinal}) using fifth-order Runge Kutta with an adaptive step size.  Because of the adaptive step size, the total number of points was not directly under our control, but rather, is determined by the desired accuracy.  For the ZTS calculations, we chose a subset of these points that were equally spaced in terms of their Cartesian distances.  When exactly equally spaced points were not available, we linearly interpolated between the two points nearest to the desired spacing.

For the third reaction, HONO elimination from dimethylnitramine, we were unable to converge a ZTS calculation from any initial estimate, likely due to the flatness of the energy surface.  We were, however, able to identify a transition state through a combination of potential energy surface scans and optimizations.  From the transition state, we generated a reaction pathway with an IRC calculation.  We compared the energy along the IRC path to the energy along 110 point paths computed with each of the four aforementioned methods,  as well as the LST method (Fig.~\ref{FigStates}).  Additionally, we used the midpoint of each path as an initial guess for a saddle-point optimization to find the transition state.  The number of geometry and SCF cycles required for each transition state calculation, along with the RMS differences between the initial estimates and the optimized transition state are given in Table~\ref{TableCompare2}.  Additional details about the results from each calculation are below.

\begin{table}[t]
		 \caption{Comparison of number of geometry and SCF cycles required for ZTS calculations starting from 11 point paths computed with various methods.}
		\centering
		\setlength{\tabcolsep}{1.5pt} 
   		\begin{ruledtabular}
			\begin{tabular}{cccc|ccc}
			&\multicolumn{3}{c}{Ethane methyl rotation}&\multicolumn{3}{c}{HCN$\to$HNC}\\
			&Geometry & SCF & Step size& Geometry & SCF & Step size\\
			\hline
			Velocity&3&173&Any&21&684&0.5\\
			Geodesic&14&440&1.0&18&678&0.1\\
			IDPP&19&529&0.5&24& 787&1.0\\
			Cartesian&67&2086&1.0&26&1174&0.5\\
			\end{tabular}
			\label{TableCompare}
   		\end{ruledtabular}
\end{table}

\subsection{Methyl rotation in ethane}
	\begin{figure}[t]
	\includegraphics[width=\columnwidth]{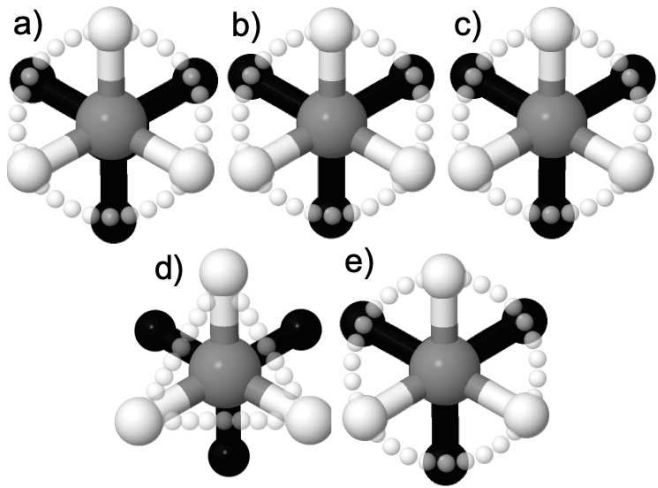}
	\caption{The estimated pathway for a 120\spt{o} rotation of a methyl group in ethane as computed with a) our velocity-based method b) the geodesic method c) the IDPP method d) linear Cartesian interpolation e) the ZTS optimized path.  The opposing methyl group is colored black for clarity.}
	\label{FigEthane}
	\end{figure}	

The endpoints of the pathway for methane rotation in ethane were optimized using the B3LYP functional \cite{Lee1988,Becke1993,Stephens1994} and 6-311++g** basis set.  These endpoints were used to generate estimated pathways with the four methods that we tested, which were then refined with a ZTS calculation using fixed endpoints.  The four estimated pathways and the final refined pathway are shown in Fig.~\ref{FigEthane}.  Atoms in the opposing, stationary methyl group are colored black for clarity, and hydrogens at intermediate points along the pathway are displayed as small, translucent spheres.

All of the estimated pathways, with the exception of the one generated from linear Cartesian interpolation, look remarkably similar, despite the drastically different number of geometry and SCF cycles required to refine them.  Our velocity-based approach significantly outperformed all of the other methods, requiring only 3 geometry cycles and 173 SCF cycles to converge.  Additionally, the optimal performance of the ZTS calculation when seeded with our estimate was independent of the step size, which was not true for the other methods we tested.

The geodesic and IDPP methods both performed better that linear Cartesian interpolation, as expected, but significantly worse than our method.  The second best performance came from the geodesic method, which required 14 geometry and 440 SCF cycles to converge.  Note that a similar test of the IDPP method for this same problem reduced the number of geometry cycles for a NEB calculation from 72, with Cartesian interpolation, to 28.\cite{Smidstrup2014} This is similar to the reduction we saw from 67 to 19.

\subsection{HCN$\to$HNC tautomerization}
\label{SecHCN}

The endpoints of the HCN$\to$HNC reaction pathway are both linear molecules.  Therefore, due to symmetry, displacement of an atom along any direction orthogonal to the axis of the molecule is equivalent.  Even after translational and rotational degrees of freedom are removed, displacements in the positive and negative directions on the axis orthogonal to the molecule are still equivalent.  This has the effect of making the metric tensor singular at the endpoints.

\begin{figure}[t]
	\includegraphics[width=\columnwidth]{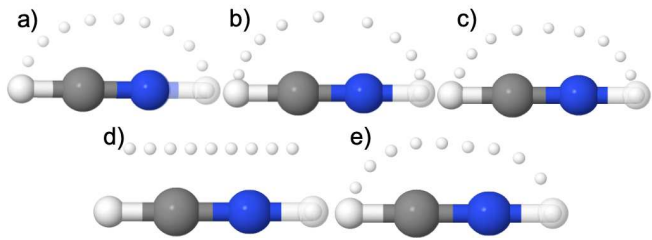}
	\caption{The estimated pathway for HCN$\to$HNC tautomerization as computed with a) our velocity-based method b) the geodesic method c) the IDPP method d) linear Cartesian interpolation e) the ZTS optimized path.  Displacements were added to the Cartesian path to avoid atomic intersections.}
	\label{FigHCN}
	\end{figure}

Any reaction pathway estimation method must artificially break this symmetry somehow.  For the path generated through linear Cartesian interpolation, we simply displaced the hydrogen atom by 1~\AA along the $Y$ axis at the nine intermediate points.  This path was used as a seed for the geodesic and IDPP methods.  In our method, we used the technique described in Sec.~\ref{SecCoordSing} for dealing with singularities at high symmetry pointes.

The endpoints of the pathway were optimized using the B3LYP functional and 6-311++g** basis set, and all estimated pathways were refined with a ZTS calculation while keeping the endpoints fixed.  All estimates performed fairly well in this problem, and there was not a big gap between the number of geometry cycles required by the best-performing method, the geodesic method, and the worst, Cartesian interpolation.  The gap in the number of SCF cycles, however, was much larger.

Our method performed second-best in terms of both the number of geometry and SCF cycles needed to achieve convergence.  Its performance was nearly identical to the geodesic method, requiring only an additional 3 geometry cycles and 6 SCF cycles.  The IDPP method, on the other hand, while it required only 6 more geometry cycles than the geodesic method, needed an additional 109 SCF cycles.  Once again, Cartesian interpolation performed significantly worse than any of the other methods.

In Fig.~\ref{FigHCN}, the pathways computed with all four estimation methods are shown, along with the ZTS refined pathway.  The pathway generated from our method looks similar to the IDPP pathway, but the geodesic pathway appears to have a slightly higher arc and the middle point is not quite centered between the carbon and nitrogen.  This latter feature is somewhat reminiscent of the ZTS path, which has a similar asymmetry of the middle point and may explain why it converged faster than any of the other paths.

\subsection{Transition state for HONO elimination from dimethylnitramine}
\label{SubsecHONO}

	\begin{figure*}[t]
	\includegraphics[width=2\columnwidth]{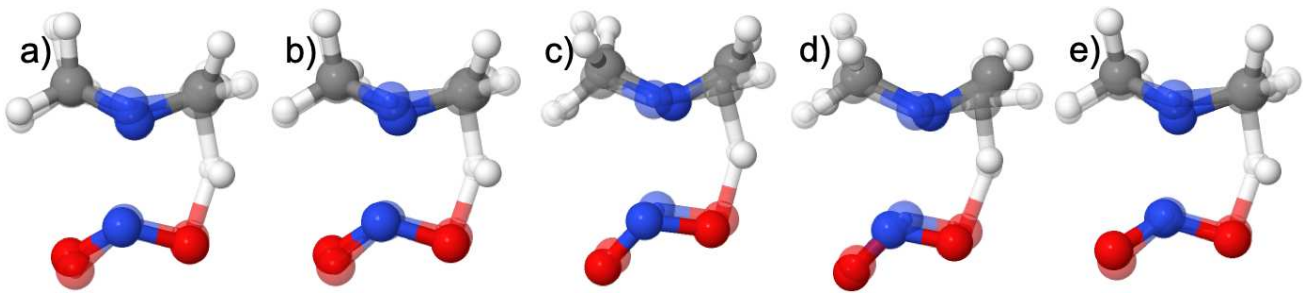}
	\caption{Estimated transition states for HONO elimination from dimethylnitramine using a) our velocity based approach b) the geodesic method c) IDPP d) linear Cartesian interpolation e) LST.  The true transition state is translucently overlaid on the guess in each image.}
	\label{FigStates}
	\end{figure*}
	
	\begin{figure}[t]
	\includegraphics[width=\columnwidth]{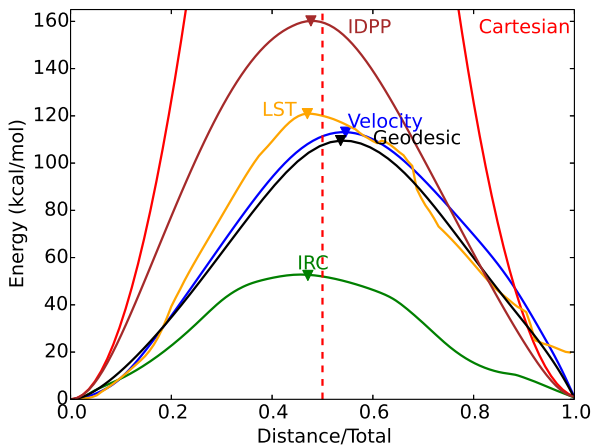}
	\caption{Energy along 110 point paths computed using interpolation of the Cartesian coordinates (red), LST (orange), IDPP (brown), the geodesic method (black), our velocity-based method (blue), and the IRC path (green).  The vertical dashed line marks the midpoint of the paths and triangular markers indicate the peaks in energy.  The peak of the Cartesian path at 676.8~kcal/mol is not included in the plot for scaling purposes.}
	\label{FigPlot}
	\end{figure}

	The HONO elimination reaction in dimethylnitramine is deceptively complex, although the molecule itself is very simple.  Nitramine compounds are essential ingredients in many propellants and explosives, and dimethylnitramine is the simplest member of the nitramine series.  It can decompose via N-N and C-H bond scission according to
	\begin{reaction}
	    \mathrm{C_2H_6N_2O_2\to CH_3NCH_2 + HONO}.
	    \label{Reaction}
	\end{reaction}
For the purposes of finding a reaction pathway between two energy minima, the terminal point of this reaction is a stable, noncovalently bonded structure consisting of CH\sbt{3}NCH\sbt{2} and HONO.
	
	The initial and final states of this reaction were optimized with the M06 functional \cite{Zhao2008} and 6-311++G(3df,3pd) basis set in Gaussian09,\cite{g09} and were used to generate 110 point paths with our velocity-based method, the geodesic method, IDPP, and LST.  The geodesic pathway was computed using the preferred internal coordinates of Zhu et al. \cite{Zhu2019} (with $\sigma=0$), and the LST pathway was computed using these same coordinates.  The energy along each of these pathways, as well as the true MEP as determined by an IRC calculation, is plotted in Fig.~\ref{FigPlot}.  Energies are given in kcal/mol relative to the reactants, and triangular markers indicate the location of their peaks.
	
	The horizontal axis in our method and the geodesic method measures distance through internal coordinates traversed at a given point along the reaction path as a fraction of the total path length.  The IRC path was also plotted in the same manner, using the same internal coordinates as the geodesic method.  For the LST and Cartesian paths, the horizontal axis is the parameter $t$, which defines the proportion of mixing between initial and final coordinates.  For the IDPP path, it is the Cartesian distance.  A vertical dashed line is placed at the midpoint, where a transition state estimate was obtained from each path.
	
	In Figure~\ref{FigStates}, the transition states estimated from each method are overlaid in 3D with the true, optimized transition state.  The true transition state is depicted with translucent atoms and bonds, for clarity.  The structures are aligned so that their centers of mass are at the origin and the RMS distance between their nuclei is minimized.

	
		\begin{table}[t]
		 \caption{Number of geometry and SCF cycles required for transition state calculations for HONO elimination from dimenthylnitramine, starting from the midpoint of 110 point paths computed with each method.  The RMS difference between the true transition state and this estimate is given in the first column.}
		\centering
   		\begin{ruledtabular}
			\begin{tabular}{cccc}
			&RMS diff. (\AA)& Geometry & SCF\\
			\hline
			Velocity&0.342 &27&348\\
			Geodesic&0.349 &41&556\\
			IDPP&0.538 & 30 & 381\\
			LST&0.396&31&396\\
			Cartesian&0.570 & 73&892\\
			\end{tabular}
			\label{TableCompare2}
   		\end{ruledtabular}
\end{table}
	
	The peak energy of the linear Cartesian path is at 676.8~kcal/mol above the reactants, which is significantly higher than the peak of the IRC path, at 52.8~kcal/mol.  This is because rather than causing one of the methyl groups to rotate as it gives up its hydrogen, linear interpolation shortens and then stretches the C-H bond-lengths, as can be seen in Fig.~\ref{FigStates} d). The RMS difference between the structure at the midpoint of this path and the true transition state structure is 0.570~\AA.  The saddle-point optimization we started from this structure eventually converged to the correct transition state after 73 geometry cycles and 892 SCF cycles.
	
	The LST path peaks at a much lower energy, at 120.9~kcal/mol, and the RMS difference between the midpoint geometry and the true transition state was 0.396~\AA.  However, we were unable to generate a continuous LST path that connected the reactant and product geometries.  By requiring both that the path is continuous and that it satisfies the LST equations at every point, the pathway does not end at an energy minimum and is geometrically quite different from the desired state.  In fact, this point, shown in Fig.~\ref{FigLST}, is approximately 20~kcal/mol above the energy of the products.  The estimate for the transition state from this method converged to the correct geometry after 31 optimization cycles and 396 SCF cycles.
	
	The midpoint of the IDPP path, despite having a relatively large RMS difference of 0.538~\AA from the optimized transition state, and a relatively high peak energy of approximately 160~kcal/mol above the reactants, converged in only 30 geometry cycles and 381 SCF cycles.  We also had a great deal of difficulty converging the IDPP path, although we were eventually able to get it to satisfy the same convergence criteria that we used in our other calculation, although only after modifying the convergence routine.  Although the IDPP method is not intended to produce transition state estimates, its performance in terms of the geometry and SCF convergence was second only to our method in this test.
	
	The geodesic path peaked at 109.5~kcal/mol and the RMS difference between the midpoint of the path and the true transition state was 0.349~\AA.  An optimization starting from this midpoint converged to the transition state after 41 geometry cycles and 556 SCF cycles.  Although these results come from a 110 point path, to test the efficiency of the algorithm, we created three different geodesic paths, with 20, 35, and 110 points.  For each path, we used two different optimization methods starting from the path computed by Cartesian interpolation.  First, we used the method described by Zhu et al. \cite{Zhu2019} where each point along the path is individually optimized, sweeping back and forth until the points stop moving.  Second, we used a Broyden-Fletcher-Goldfarb-Shanno (BFGS) optimization scheme that updated all intermediate points on the path at once.
	
	The first optimization method required 26~s to compute the 20 point path, 220~s for the 35 point path, and 5.12~hours for the 110 point path.  These times scale slightly worse than cubically with the number of points.  Our BFGS implementation, by contrast, computed the 20 point path in 3~seconds, the 35 point path in 15~s, and the 110 point path in 476~s.  The ratio of these times are almost exactly the cube of the ratio of the number of points, indicating a computational time that scales as $O(P^3N^3)$, which is reasonable given the fact that the size of the Hessian is $PN\times PN$, and (despite the fact that the BFGS method does not actually involve inverting the Hessian) the time required to solve a linear system is cubic in the number of dimensions.
	
	However, because each point only couples to its nearest neighbors, the Hessian has a block-tridiagonal structure consisting of $P$ overlapping rows of three blocks, each of size $(3N-6)\times(3N-6)$.  Linear systems of this form can be solved in $O(PN^3)$ operations,\cite{Fourer1984}, and a sparse Quasi-Newton algorithm that exploits this structure, such as the one described by Yamashita \cite{Yamashita2008}, could likely make the computation time linear in $P$.
	
	Finally, our velocity-based method performed better than all of the others by every metric, except for the peak energy, which at 114.5~kcal/mol, was slightly higher than the geodesic method.  The RMS difference between our midpoint and the true transition state was the lowest of any method at 0.342~\AA, and the saddle-point optimization reached the true transition state in just 27 geometry cycles and 348 SCF cycles.  Unlike LST the pathway generated by our method was continuous and terminated at the correct geometry.  The computational time of 0.07~s for the tightest integration criteria was two orders of magnitude smaller than time required to compute even the smallest path with the geodesic method using our fast BFGS optimization.  This is likely because our method requires solving a linear system involving an $N\times N$ symmetric matrix (the metric tensor) exactly once at each point, and thus scales linearly with the number of integrations points, as $O(PN^3)$.
	
		We tested various accuracies of the Runge-Kutta stepper routine in our method, between 10\spt{-8} and 10\spt{-4}, without noticeable degradation of the pathway.  The calculation was terminated when the RMS difference between initial and final geometries was less than 10\spt{-4}~\AA.  For loosest integration accuracy, the processor time to compute the reaction pathway was approximately 0.02~s, and for the tightest, it was approximately 0.07~s.

\subsection{The linear synchronous transit pathway}
\label{SecLSTCompare}
\begin{figure}[t]
\includegraphics[width=0.45\columnwidth]{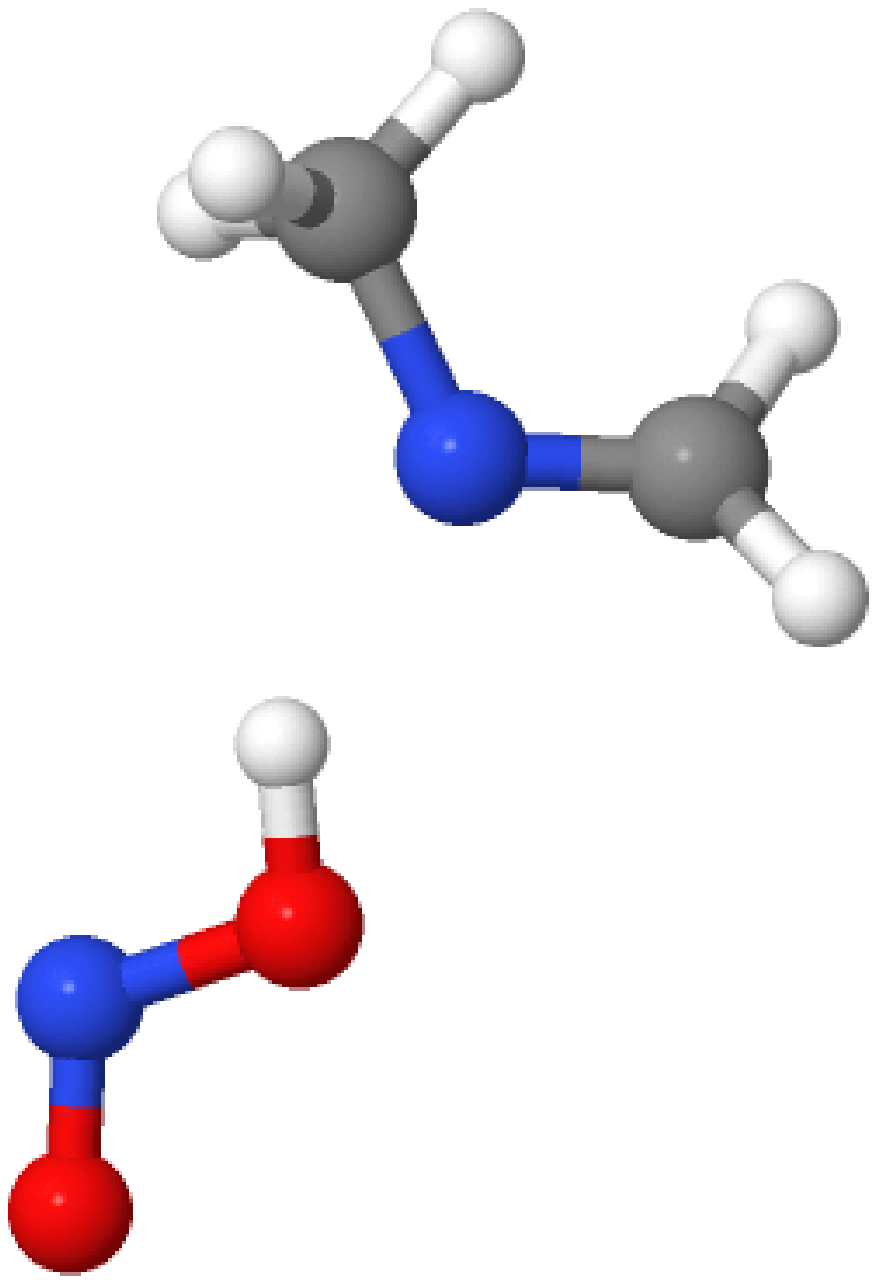}
\includegraphics[width=0.49\columnwidth]{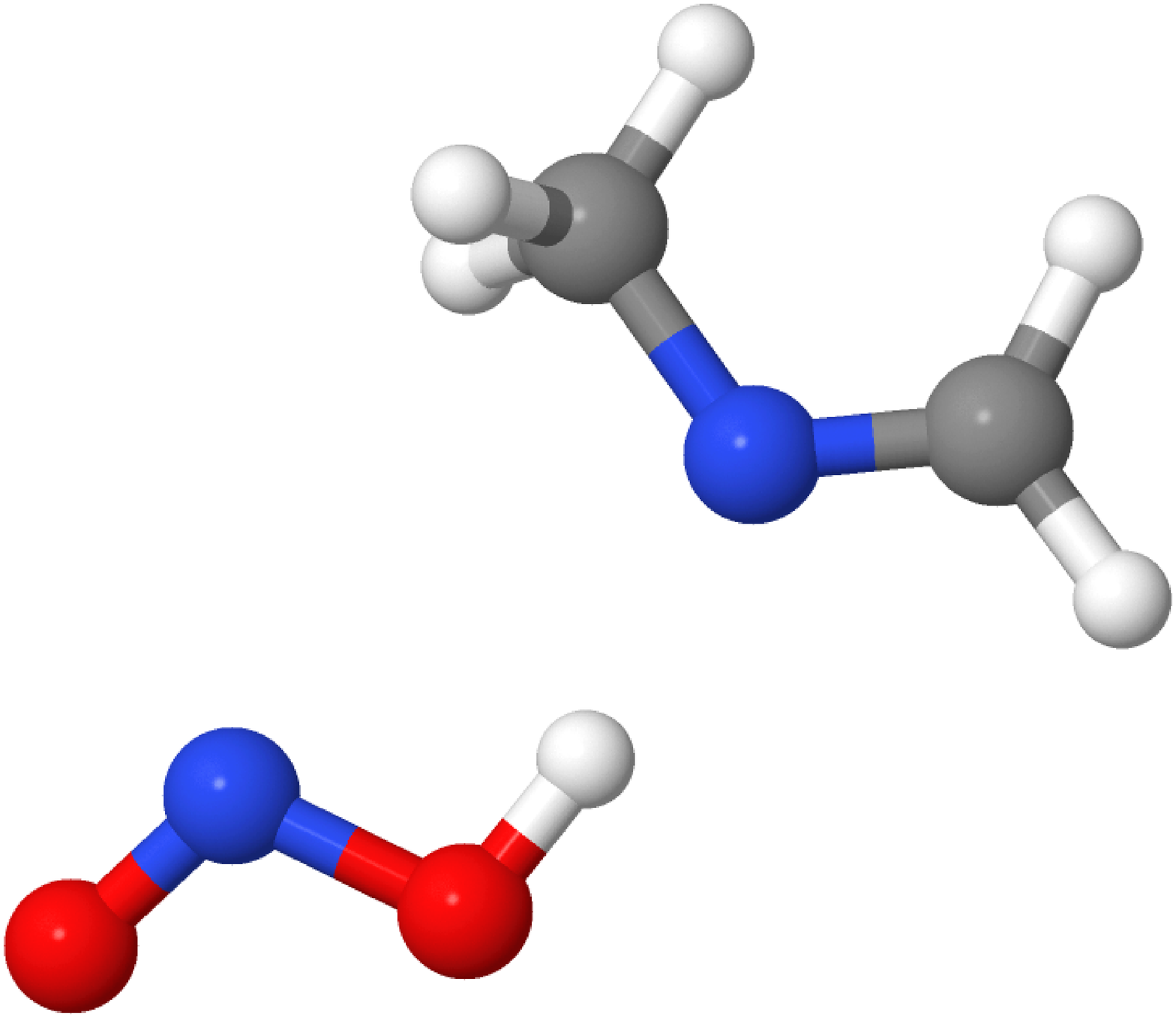}
\caption{The terminal point when attempting to generate a continuous LST path (left) versus the correct terminal point (right) for the HONO elimination from dimethylnitramine.}
\label{FigLST}
\end{figure}

Frequently, LST does not produce a continuous pathway connecting reactants and products.  However, it is well understood that it is usually possible to create a continuous path on the interval $0\leq t\leq 1$ that satisfies the LST equations at every point, although it is not guaranteed to end at the proper terminal point.  We describe our procedure for generating such a path here

To generate a continuous LST path, we started from Eq.~(\ref{EqLST}) and produced a first-order differential equation by taking its time derivative
\begin{equation}
    \frac{d}{dt}\frac{\partial}{\partial x^j} \sum_i\left(q^i-(1-t)q^{i0}-tq^{i1}\right)^2 = 0,
\end{equation}
which yields
\begin{equation}
\begin{split}
&\frac{\partial q^i}{\partial x^j}\left(\frac{\partial q_i}{\partial x^k}\frac{dx^k}{dt} + q_{i0} - q_{i1}\right)\\
&+\left(q^i-(1-t)q^{i0}-tq^{i1}\right)\frac{\partial^2 q_i}{\partial x^k\partial x^j}\frac{dx^k}{dt}=0,
\end{split}
\end{equation}
where upper and lower indices are equivalent, because the metric is Euclidean.

Starting from the geometry of the reactants and integrating this equation will produce a continuous path from reactants to products that satisfies the LST equations so long as such a path exists.  In practice, we found that this happens very infrequently.  In Fig~\ref{FigLST}, we show the terminal point of the resulting path for HONO elimination in DMNT.

The Runge-Kutta integrator runs into some difficulty near $t=0.72$ and took over a minute of runtime to move past this point.  In the computed pathway, this corresponds to a quick leap in the positions of the hydrogens on the methyl group.  After this point, the HONO group undergoes some unusually jerky translations and rotations until finally reaching the point depicted in Fig.~\ref{FigLST}.

\section{Conclusions}
	
	The velocity-based method that we developed provides a good estimate of MEPs and transition states for refinement with higher-level electronic structure theory.  Unlike many other pathway estimation techniques, it does not need to be seeded with an even simpler estimate.  Therefore, no additional work is required to avoid atomic intersections and unlike variational approaches, it cannot converge to an undesirable local minimum.  It is based on the premise that a straight line through some set of generalized coordinates can approximate the minimum energy pathway between two states.  Although this path is usually not realizable in terms of actual atomic positions, we can define a velocity vector field in generalized coordinates that takes the reactant geometry to the product geometry.  This vector field can then be projected from the high-dimensional manifold of generalized coordinates into the Cartesian manifold with translational and rotational degrees of freedom removed.  We do this by taking the corresponding velocity 1-form in internal coordinates, pulling it back into Cartesian coordinates, and acting on it with the inverse metric tensor to produce a vector.
	
	We tested our method against the geodesic method of Zhu et al., IDPP, and linear Cartesian interpolation for estimating MEPs, and additionally against LST for estimating transition states.  The estimated pathway generated by our method for a 120\spt{o} rotation of a methyl group in ethane converged to the true MEP, as determined by a ZTS calculation, significantly faster than any other method.  Convergence was achieved from our path in only 3 geometry and 173 SCF cycles, whereas the second best-performing method required 14 geometry and 440 SCF cycles, and Cartesian interpolation required 67 geometry and 2086 SCF cycles.  In our second test, HCN$\to$HNC tautomerization, the geodesic method slightly outperformed ours, although our method converged in only 6 extra SCF cycles, as opposed to 109 extra SCF cycles for the next best-performing method, IDPP.
	
	In finding the transition state for HONO elimination from dimethylnitramine, our method once again led to the fastest convergence of any of the methods we tested.  Our midpoint was also the closest geometrically to the true transition state.  While the midpoint of the geodesic method was nearly as close geometrically, and slightly closer energetically, it required a surprising number of additional geometry and SCF cycles to reach convergence.  Cartesian interpolation performed particularly badly in this system because it caused the shortening, rather than rotation, of C-H bonds in one of the methyl groups.  Because of this, the peak energy of the Cartesian path was approximately 630~kcal/mol above the true transition state and 562~kcal/mol above the peak of our path.
	  
	We were able to compute an estimate for the 110 point path for HONO elimination from dimethylnitramine using our method in 0.07~s, which was several orders of magnitude faster than the geodesic method (and although we did not do any detailed analysis, it was also several orders of magnitude faster than our implementation of IDPP and LST).  It should be pointed out, however, that the times required to generate estimated paths with all of the methods we tested were very short compared to the times required to refine them with electronic structure theory.  Still, it is worth noting that the computational time of our method scales linearly with the number of points on the path, and the implementation of the geodesic method described by Zhu et al. appears to scale cubically, although we believe that it should be possible to create a linear-scaling version of their algorithm as well.
	
	Our method is capable of quickly generating estimates of reaction pathways and transition states that can be optimized via electronic structure theory in relatively few geometry and SCF cycles, and appears in several cases to outperform other existing methods.  We expect, therefore, that this technique will be a valuable addition to the arsenal of reaction pathway estimation tools.
	
		
	\begin{acknowledgments}
		This work was supported by the Office of Naval Research (ONR) directly and through the Naval Research Laboratory.  We thank Igor V. Schweigert for providing transition state and IRC calculations for dimethylnitramine.
	\end{acknowledgments}

\bibliography{citations}
\end{document}